\documentclass[sigconf]{acmart}

\copyrightyear{2025}
\acmYear{2025}
\setcopyright{rightsretained}
\acmConference[CF Companion '25]{22nd ACM International Conference on
Computing Frontiers}{May 28--30, 2025}{Cagliari, Italy}
\acmBooktitle{22nd ACM International Conference on Computing Frontiers (CF
Companion '25), May 28--30, 2025, Cagliari,
Italy}

\usepackage[per-mode=symbol,retain-explicit-plus,separate-uncertainty=true]{siunitx}
\usepackage{url}
\usepackage{breakurl}
\usepackage{graphicx, nicefrac}
\usepackage{lipsum}
\usepackage{hyperref}
\usepackage{dblfloatfix}    
\usepackage{enumitem}
\usepackage{caption}
\usepackage{natbib}
\usepackage{multirow}
\usepackage{graphicx}
\usepackage{colortbl}

\usepackage[acronym, toc,numberedsection = nolabel, section = part]{glossaries}
\usepackage{graphicx}
\usepackage{booktabs}
\usepackage{tabularx}

\makenoidxglossaries
\newacronym{alap}{ALAP}{As-Late-As-Possible}
\newacronym{asap}{ASAP}{As-Soon-As-Possible}
\newacronym{alu}{ALU}{Arithmetic-Logic Unit}
\newacronym{asic}{ASIC}{Application Specific Integrated Circuit}
\newacronym{cgra}{CGRA}{Coarse-Grain Reconfigurable Array}
\newacronym{cnf}{CNF}{Conjunctive Normal Form}
\newacronym{cnn}{CNN}{Convolutional Neural Network}
\newacronym{cil}{CIL}{Compute-Intensive Loop}
\newacronym{dfg}{DFG}{Data Flow Graph}
\newacronym{dma}{DMA}{Direct Memory Access}
\newacronym{fpga}{FPGA}{Field Programmable Gate Array}
\newacronym{ims}{IMS}{Iterative Modulo Scheduling}
\newacronym{isa}{ISA}{Instruction Set Architecture}
\newacronym{ilp}{ILP}{Integer Linear Programming}
\newacronym{ir}{IR}{Intermediate Representation} 
\newacronym{ii}{II}{Iteration Interval}
\newacronym{im2col}{Im2col}{Image-to-Column}
\newacronym{ip}{IP}{Input-Channel Parallelism}
\newacronym{kms}{KMS}{Kernel Mobility Schedule}
\newacronym{mii}{mII}{minimum II}
\newacronym{ms}{MS}{Mobility Schedule}
\newacronym{mac}{MAC}{Multiply-and-Accumulate}
\newacronym{op}{OP}{Output-Channel Parallelism}
\newacronym{os}{OS}{Output Stationary}
\newacronym{pc}{PC}{Program Counter}
\newacronym{pe}{PE}{Processing Element}
\newacronym{ra}{RA}{Register Allocation}
\newacronym{rf}{RF}{Register File}
\newacronym{rtl}{RTL}{Register Transfer Level}
\newacronym{sat}{SAT}{satisfiability}
\newacronym{soa}{SoA}{State of the Art}
\newacronym{u}{U}{Utilization}
\newacronym{ws}{WS}{Weight Stationary}
\newacronym{wp}{WP}{Weight Parallelism}

\glsdisablehyper

\copyrightyear{2025}
\acmYear{2025}
\setcopyright{rightsretained}
\acmConference[CF Companion '25]{22nd ACM International Conference on
Computing Frontiers}{May 28--30, 2025}{Cagliari, Italy}
\acmBooktitle{22nd ACM International Conference on Computing Frontiers (CF
Companion '25), May 28--30, 2025, Cagliari,
Italy}\acmDOI{10.1145/3706594.3726977}
\acmISBN{979-8-4007-1393-4/2025/05}

\begin{document}

\title{A flexible framework for early power and timing comparison of time-multiplexed CGRA kernel executions}

\author{Maxime Henri Aspros}
\affiliation{%
  \institution{Polytechnique Montreal}
  \city{Montreal}
  \country{Canada}}
\email{maxime.aspros@polymtl.ca}

\author{Juan Sapriza}
\affiliation{%
  \institution{École Polytechnique Fédérale de
 Lausanne}
  \city{Lausanne}
  \country{Switzerland}}
\email{juan.sapriza@epfl.ch}

\author{Giovanni Ansaloni}
\affiliation{%
  \institution{École Polytechnique Fédérale de
 Lausanne}
  \city{Lausanne}
  \country{Switzerland}}
\email{giovanni.ansaloni@epfl.ch}

\author{David Atienza}
\affiliation{%
  \institution{École Polytechnique Fédérale de
 Lausanne}
  \city{Lausanne}
  \country{Switzerland}}
\email{david.atienza@epfl.ch}

\settopmatter{printacmref=true}

\renewcommand{\shortauthors}{Aspros, et al.}

\begin{abstract}

At the intersection between traditional CPU architectures and more specialized options such as FPGAs or ASICs lies the family of reconfigurable hardware architectures, termed Coarse-Grained Reconfigurable Arrays (CGRAs). CGRAs are composed of a 2-D array of processing elements (PE), tightly integrated with each other, each capable of performing arithmetic and logic operations.
The vast design space of CGRA implementations poses a challenge, which calls for fast exploration tools to prune it in advance of time-consuming syntheses.
The proposed tool aims to simplify this process by simulating kernel execution and providing a characterization framework.
The estimator returns energy and latency values otherwise only available through a time-consuming post-synthesis simulation, allowing for instantaneous comparative analysis between different kernels and hardware configurations.

\end{abstract}

\ccsdesc[100]{Computer systems organization~Embedded software}

\keywords{CGRA, Design Space Exploration, Simulator}

\maketitle

\section{Introduction}
\glsresetall{}
While CPU-based architectures have long dominated edge computing, their efficiency has been shown to be short when executing tasks that could benefit from parallelization, such as those widely present in edge-AI applications. 
Several efforts have been made to provide parallelization and task-specific capabilities to MCU, such as clusters of CPUs~\cite{flamand2018gap, garofalo2020pulp} and task-specific accelerators \cite{GreenWaves2024}. 
The first provide great flexibility at the cost of area and power overhead. The latter offer high efficiency at the expense of versatility to accelerate other tasks that might be required during the application. Coarse-Grained Reconfigurable Arrays (CGRAs)~\cite{podobas2020survey} have gained attention as meet-in-the-middle solutions that provide parallelization through an array of simple Processing Elements (PEs), which are capable of executing a limited set of arithmetic and logic operations, providing flexibility at a low overhead.   
CGRAs are typically classified into two categories based on how they manage processing capabilities and scheduling. Spatial models, in which the configuration of each PE remains unchanged until the execution of a kernel finishes, only requiring a static compilation \cite{vazquez2024strelastreamingelasticcgra}. CGRAs that are time-multiplexed differ in that they support different PE configurations (CGRA instructions) during execution. 

\begin{figure}[t]
\vspace{0.5cm}
    \centering
        \includegraphics[width=1\linewidth]{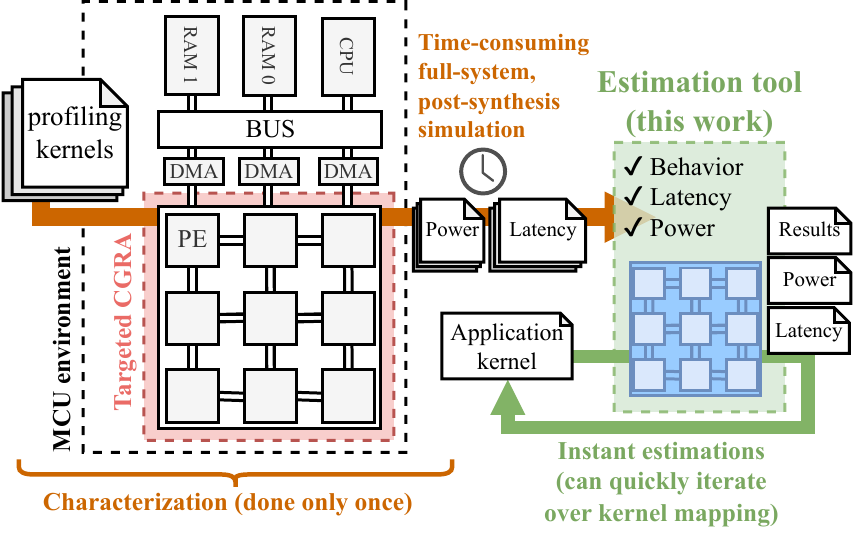}
    \caption{Workflow of the estimation tool. Profiling kernels are used to obtain a characterization in power and latency of the targeted CGRA (as part of a broader MCU environment). These one-time-results are used to create a model over which application kernels can be simulated to obtain accurate and instant power and timing results.}
    \label{fig:intro}
\vspace{-10pt}
\end{figure}

In time-multiplexed CGRAs such as \cite{alvarez2023open, Duch227874, Das2017, Bandara2022}, compute-intensive loops (kernels) are mapped across a range of PEs and time (instructions). At each iteration, the CGRA executes one instruction, which consists of a unique operation for each of its PEs. The execution of the whole kernel depends on the coordination of neighboring PEs, and therefore they all share a common program counter (PC) and will advance to the next instruction simultaneously once all PEs have finished. Each PE can execute an operation from the CGRA's ISA and take arguments either from immediate values, their own set of registers, or neighboring PEs. CGRAs are usually not instantiated standalone, but rather are used as an accelerator of a broader system encompassing CPU, memory and other peripherals. If the CGRA does not have its own dedicated memory, it must share the access to memory with other elements in the system. 
This complex structure makes mapping kernels to time-multiplexed CGRAs a challenge due to spatial, temporal, and system-dependent considerations and the need to manage various levels of freedom.
Large efforts have been made to develop compilers that facilitate this task~\cite{tirelli2024sat, wang2025mlir, mei2003exploiting, wijerathne2021himap}, but they still fall short of considering the effect of the whole system on the CGRA's execution. 
Additionally, simulating the execution of a kernel usually requires a time-consuming RTL simulation, and obtaining power estimates even requires slower post-synthesis simulations.
Previous estimators include CGRA-EAM \cite{Wijtvliet2021}, which proposes an energy and area estimation framework for space-multiplexed CGRAs. However, this model is data agnostic and therefore cannot leverage run-time information, instead relying on an execution trace.

Recognizing this challenge, this work streamlines the analysis process by introducing a behavioral simulation tool, combined with a customizable estimator. The estimator’s workflow allows for an incremental adjustment of precision, according to the number of accounted non-idealities. 
For example, an initial run can focus mainly on latency, with a power consumption that scales linearly with the duration of an instruction. The precision of the estimation can then be further enhanced with specifications such as the chosen bus type or even certain properties at the PE level, including the datapath switch and register choice. 

\begin{table}[t!]
    \centering
    \caption{Considered non-idealities and corresponding cases}
    \includegraphics[width=0.9\linewidth]
    {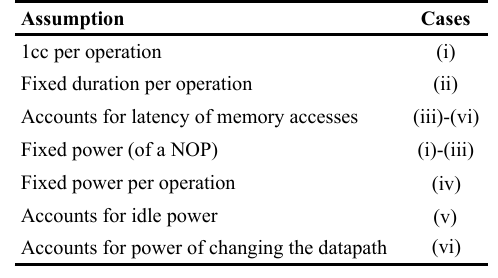}
    \label{tab:nonideality}
\vspace{-15pt}
\end{table}

\section{Validation}\label{sec:methodology}

The estimation tool presented in this work (green in \autoref{fig:intro}) is open-source\footnote{Available at: \href{github.com/esl-epfl/ESL-CGRA-simulator}{github.com/esl-epfl/ESL-CGRA-simulator}}, and is entirely Python-based. A model of the target CGRA and its ISA is built as a set of modular Python functions (blue in \autoref{fig:intro}), which allows for behavioral simulation. The state of each PE and its internal registers in each cycle can be observed to debug the application kernel. To include power and latency estimations, a characterization model needs to be provided. The target CGRA is to be profiled using custom kernels to obtain power and latency values for different non-idealities that want to be modeled (red in \autoref{fig:intro}). Using the characterization, our tool is able to estimate the power and latency for each PE and cycle, providing a full profiling of the application kernel without incurring in any more time-consuming post-synthesis simulations. Due to the instantaneous result, this work allows for rapid iterations over software and hardware. Once satisfied with the result, the final instructions are encoded into a bitstream to be deployed into the CGRA.  In this work, we model the OpenEdgeCGRA~\cite{alvarez2023open}, an open-hardware low-power CGRA to validate the model's performance, but handle other instruction-based CGRAs given sufficient characterization information.

\begin{figure}[t!]
\vspace{-10pt}
    \centering
    \includegraphics[width=1\linewidth]{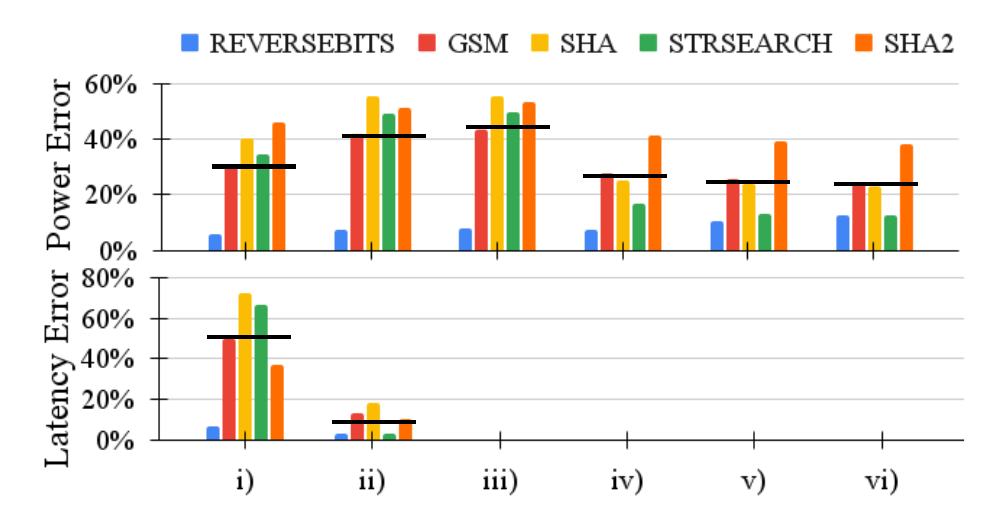}
    \caption{Impact of non-idealities on precision. Absolute error obtained from executing five benchmark kernels on the tool of this work vs results obtained from post-synthesis simulation. The average of each case is marked in black.}
    \label{fig:nonideality-graph}
    \vspace{-10pt}
\end{figure}

The quality of the estimation depends on the level of detail provided by the user in the characterization profile. A basic version of the model (case (i) from \autoref{tab:nonideality}), using a uniform latency and power value for each operation, presents a significant discrepancy between behavioral analysis and expected values. This is showcased in \autoref{fig:nonideality-graph}, comparing post-synthesis simulations of the \hbox{OpenEdgeCGRA} in TSMC 65nm LP process against the results from our tool. 
By progressively including non-idealities, the user can help strengthen the subsequent models (ii-vi). This requires writing specific test kernels to characterize the CGRA's profile. 
\autoref{tab:nonideality} lists the parameters used for the sample explorations, while \autoref{fig:nonideality-graph} illustrates the decrease in latency and power error after including each non-ideality.

The first three iterations represent the step-by-step refinement of latency estimation. As the tool incorporates additional parameters from the characterization file, precision improves. 
Step (ii) considers a unique latency value for each operation, obtained from the characterization file. In the case of \hbox{OpenEdgeCGRA}, all logic and arithmetic operations take 1 clock cycle (cc), with the exception of multiplication (3 cc) and memory accesses, which vary drastically depending on the state of the rest of the system. 
The next step (iii) considers in addition this delay incurred while waiting for memory accesses. 

The following three iterations focus on power estimation. Step (i) begins with a basic approach, applying the same consumption to all operations (in this case, the power of executing a NOP operation). 
Step (iv) considers a fixed power per operation, disregarding a consumption profile across the execution of the instruction (similar to step (ii)).
Step (v) is analogous to step (iii), as it considers the idle power consumed by a PE after executing its operation and while waiting for other PEs on the CGRA to finish their execution.
Step (vi) considers changes in the datapath, i.e. the cost of switching muxes if the operation of a PE changes from one instruction to the next, and the cost different arguments may require (e.g. distinguishes between multiplying by 0 and other values, or if the arguments are fetched from an immediate, a register or a neighbouring PE). 

Although this example uses values specific to \hbox{OpenEdgeCGRA}, the estimator remains flexible and allows multiple variations with different CGRAs. These variations can be supported by updating the characterization file with values obtained through testing or customizing the ISA by modifying the behavior of the operation functions in the Python model. 

Upon recording values taken from the test kernels, the model was validated on five kernels from the MiBench benchmark suite ~\cite{guthaus2001mibench}, obtaining the results shown in \autoref{fig:nonideality-graph}. It ultimately achieves a 22\% final error in power consumption, while the latency error decreases from to 46\% to 9\% with the first non-ideality and reaches the expected value by the third. This performance indicates that the profile accurately models latency and allows for comparative studies of power consumption. More importantly, as Section \ref{sec:results} will address, our approach handles comparative explorations of different implementations in both software and hardware.

\section{Use cases}\label{sec:results}
This section presents applications of the estimator through two studies. The first study highlights the tool's potential for software exploration, with an example featuring several mappings of a same function. We then examine a second application, this time aimed towards exploring different hardware configurations. All simulations are carried out considering the most non-idealities as in \hbox{case (vi)}.

\vspace{-5pt}
\subsection{Software exploration: Same hardware, same function, different instructions}\label{sec:mappings}
\begin{figure}[t!]
    \centering
    \includegraphics[width=1\linewidth]{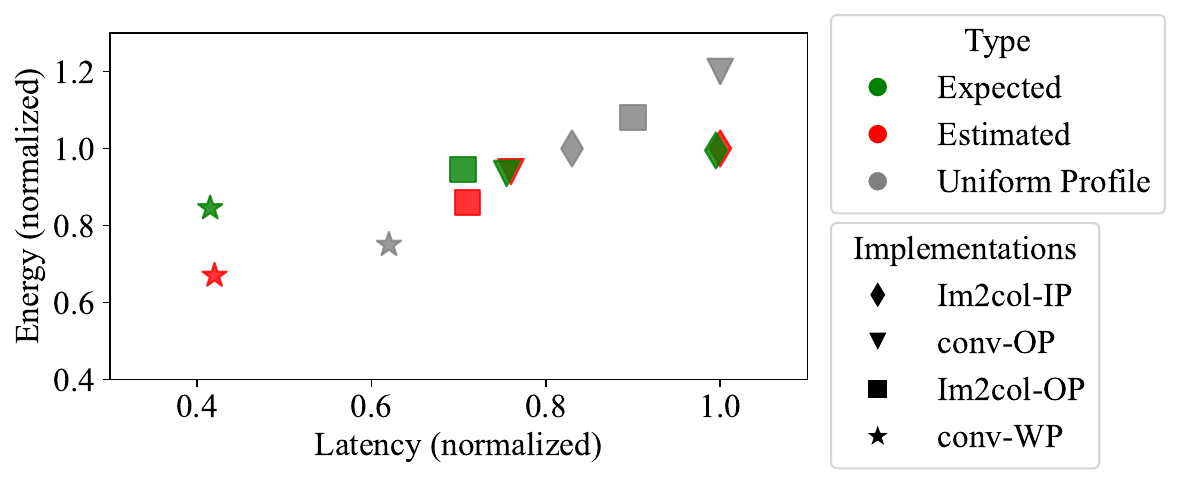}
    \caption{CGRA energy vs. latency comparison for convolution mappings. Values are normalized to the expected (post-synthesis) values for Im2col-IP.}
    \label{fig:rc-difference}
    \vspace{-10pt}
\end{figure}

\begin{figure}[t!]
    \centering
    \includegraphics[width=1\linewidth]{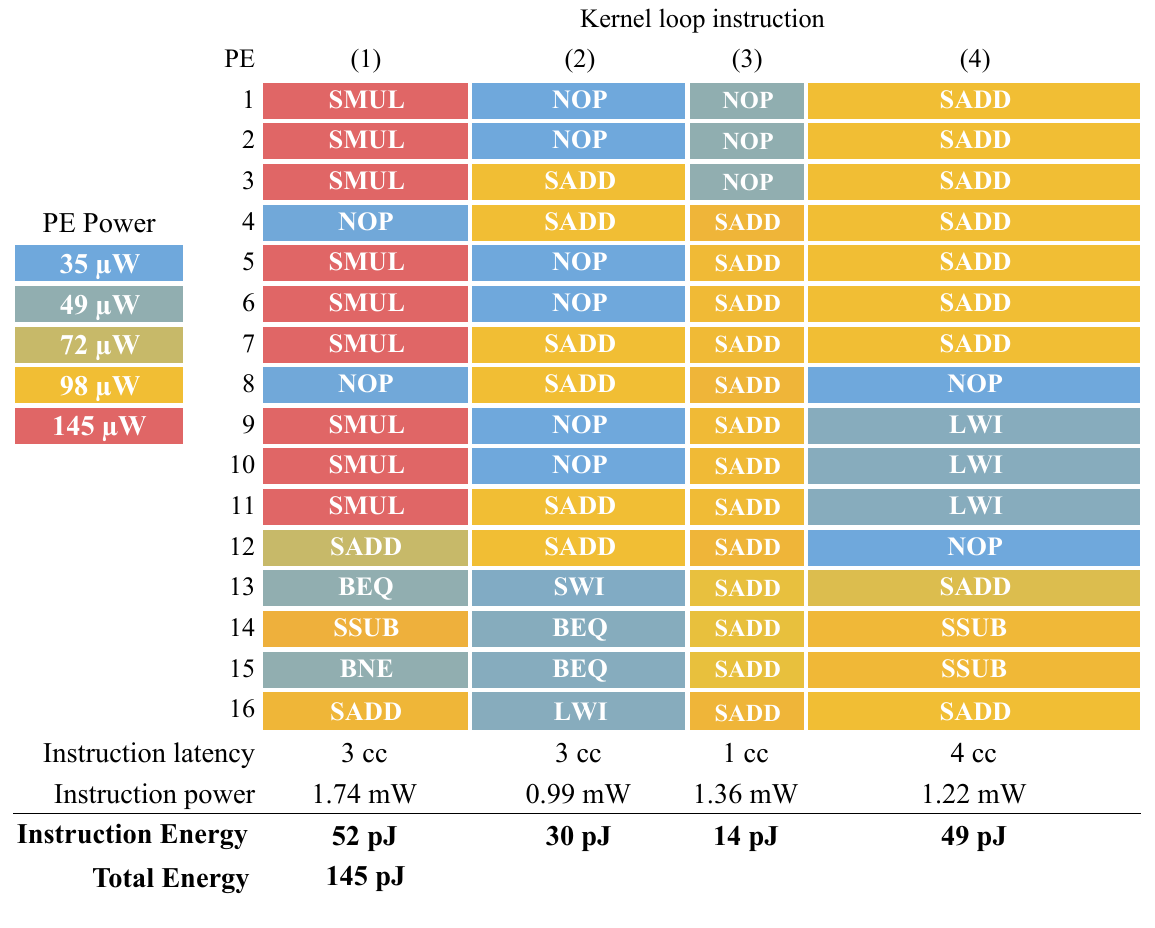}
    \caption{Power Consumption heatmap for the four instructions conforming the kernel loop of the WP convolution}
    \label{fig:instr-heatmap}
    \vspace{-10pt}
\end{figure}

Figure \ref{fig:rc-difference} presents an experiment that considers multiple convolutional mappings explored in ~\cite{Carpentieri2024}: Weight Parallelism (conv-WP), Input-Channel Parallelism (Im2col-IP), Output-Channel Parallelism (Im2col-OP), and Channel Output Parallelism (conv-OP). These various implementations use different strategies to compute a convolution and produce the same result. 
In~\cite{Carpentieri2024}, time-consuming post-synthesis simulations were run for each in order to evaluate behavior correctness and obtain latency and energy estimations to choose the most convenient approach. We replicated these results (green in \autoref{fig:rc-difference}) and compared them with the results obtained from our tool (red in \autoref{fig:rc-difference}). The close correlation between the obtained values shows how our tool can be used to choose the most convenient instruction mappings without incurring in time-consuming simulations. 
For reference, \autoref{fig:rc-difference} also shows in gray the results that would have been obtained in case (i) of \autoref{tab:nonideality}, which highlights the relevance of a proper characterization in order to draw relevant conclusions. 
In addition to reduced runtime, our tool can also offer a fine-grained breakdown of power and latency consumption. For the four instructions in the loop of the conv-WP convolution, the power and latency of each PE in a $4\times4$ \hbox{OpenEdgeCGRA} is shown in \autoref{fig:instr-heatmap}. 
For instance, it can be observed how the average power of instructions such as the NOP decreases over time as the power required during instruction decoding is much greater than the power consumed waiting for other PEs. This means that clustering time-consuming operations in a single, long instruction helps reduce energy consumption vs. having several shorter instructions. It can also be observed how the largest contribution to overall energy is not operation power, but instead operation latency. Although CGRA instruction (1) is the most power hungry due to 9 signed multiplication (SMUL) operations, its overall energy is comparable to CGRA instruction (4), where most instructions are power-light, but waiting for memory accesses (LWI) drastically increases instruction energy. 
In the following study we will compare the effect on energy of reducing the latency of SMUL and memory access instructions through architectural changes and reinforce this observation. 

\vspace{-5pt}
\subsection{Hardware exploration: Same function, same instructions, different hardware}

In this study, only the conv-WP implementation is analyzed, but different hardware architectures are varying for the CGRA and its connection to the overall MCU system. 
In the state-of-the-art, similar architectural changes, such as memory bus modifications, MAC units, and operand reuse, currently require an extensive manual configuration prior to execution. ~\cite{lee2021specializing, Carpentieri2024, heidorn2020design}.
In this scenario, the compared models contain structural modifications that involve changes in the RTL. In addition to the time required for a standard post-synthesis, these deeper changes habitually require rebuilding the model and must be entered by the user, extending the workflow even more. Using the estimator, the user can enter the changes and immediately execute the program afterwards.  

\begin{figure}[b!]
\vspace{-10pt}
    \centering
    \includegraphics[width=0.95\linewidth]{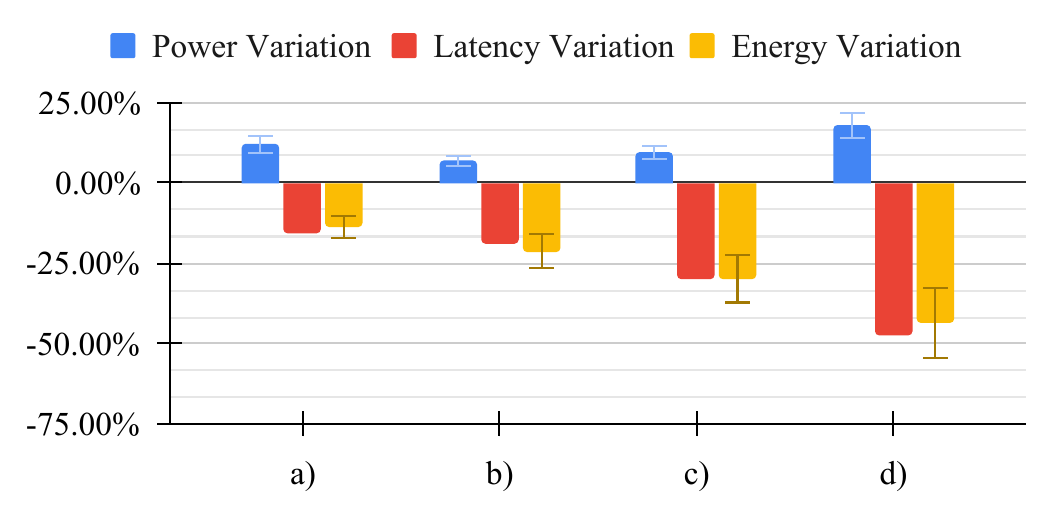}
  \caption{Hardware Topology Impact on Optimization}
    \label{fig:hw-topology}
    \vspace{-10pt}
\end{figure}

The percentage reduction in Figure \ref{fig:hw-topology} is measured relative to the baseline architecture used in previous analysis. The error bars featured in the chart account for the 22\% error obtained for the power estimation (see \autoref{fig:nonideality-graph}).
The first \hbox{modification (a)} considers a multiplication operation that completes in one cc instead of three. To adjust for this reduced delay, the operation's power is increased threefold. The second \hbox{modification (b)} changes the bus type from 1-to-M to an N-to-M, thus allowing for parallel memory accesses (given they target different memory banks).
The third \hbox{modification (c)} chooses to use an interleaved bus, while, the last \hbox{modification (d)} simulates an architecture with one DMA per PE instead of one per column, and an N-to-M bus. This bus type must be selected to obtain the gains from adding more DMA ports.

In modification (a), the accelerated multiplication naturally leads to a decrease in total latency. However, because of the updated multiplication consumption, the overall power usage increases proportionally and the energy gains are marginal. 
Conversely, all changes relative to memory accesses (b-d) significantly reduce delays. As a result, the program waits less and executes more power-intensive instructions, increasing average power consumption. These cases confirm the role of accelerated memory accesses in decreasing energy, thus aligning with the analysis in \autoref{sec:mappings}. 
Modification (d) lowers latency the most, as it can potentially remove any delay caused by multiple memory accesses in one instruction. Note that our estimation focuses only on the PE matrix, so the cost of additional DMA ports is not considered. \\ 

\begin{table}[t!]
    \centering
    \caption{Hardware topology explorations and their corresponding cases}
    \vspace{0.25cm}
    \includegraphics[width=7cm]{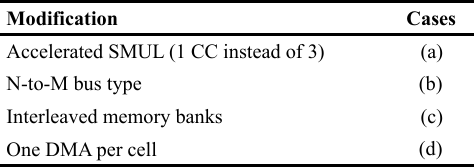}
    \label{tab:hw-cases}
    \vspace{-10pt}
\end{table}

\vspace{-10pt}
\section{Conclusions}\label{sec:conclusions}
This work has featured a new and flexible framework to estimate CGRA kernel execution times, and has illustrated the underlying factors required to build an appropriate characterization profile for multi-objective application optimization. We established how an iterative process of non-ideality inclusion maximizes the quality of the tool's performance, achieving an average error in power consumption of 22\% for the benchmark kernels with no error in latency, and a 10\% power consumption error with a 0.4\% error in latency for the convolutions.  
Our comparisons across the convolution kernels reveal how the tool characterizes mappings across both energy and latency metrics, while hardware exploration highlighted the performance gains made apparent from testing different topologies. Together, these features derived actionable insights across both co-design dimensions in an iterative process.

\vspace{-5pt}
\begin{acks}
This work was supported in part by the the Swiss NSF Edge-Companions project (GA No. 10002812); in part by the EC H2020 FVLLMONTI Project under Grant 101016776; in part by the ACCESS—AI Chip Center for Emerging Smart Systems, sponsored by InnoHK funding, Hong Kong, SAR; and in part by the Swiss State Secretariat for Education, Research, and Innovation (SERI) through the SwissChips Research Project.
\end{acks}

\vspace{-5pt}

\bibliographystyle{ACM-Reference-Format}
\bibliography{main}


\begin{thebibliography}{18}


\ifx \showCODEN    \undefined \def \showCODEN     #1{\unskip}     \fi
\ifx \showDOI      \undefined \def \showDOI       #1{#1}\fi
\ifx \showISBNx    \undefined \def \showISBNx     #1{\unskip}     \fi
\ifx \showISBNxiii \undefined \def \showISBNxiii  #1{\unskip}     \fi
\ifx \showISSN     \undefined \def \showISSN      #1{\unskip}     \fi
\ifx \showLCCN     \undefined \def \showLCCN      #1{\unskip}     \fi
\ifx \shownote     \undefined \def \shownote      #1{#1}          \fi
\ifx \showarticletitle \undefined \def \showarticletitle #1{#1}   \fi
\ifx \showURL      \undefined \def \showURL       {\relax}        \fi
\providecommand\bibfield[2]{#2}
\providecommand\bibinfo[2]{#2}
\providecommand\natexlab[1]{#1}
\providecommand\showeprint[2][]{arXiv:#2}

\bibitem[{\'A}lvarez et~al\mbox{.}(2023)]%
        {alvarez2023open}
\bibfield{author}{\bibinfo{person}{Rub{\'e}n~Rodr{\'\i}guez {\'A}lvarez} {et~al\mbox{.}}} \bibinfo{year}{2023}\natexlab{}.
\newblock \showarticletitle{{An Open-Hardware Coarse-Grained Reconfigurable Array for Edge Computing}}. In \bibinfo{booktitle}{\emph{Proc. of the 20th ACM CF}}. \bibinfo{pages}{391--392}.
\newblock


\bibitem[Bandara et~al\mbox{.}(2022)]%
        {Bandara2022}
\bibfield{author}{\bibinfo{person}{Thilini~Kaushalya Bandara}, \bibinfo{person}{Dhananjaya Wijerathne}, \bibinfo{person}{Tulika Mitra}, {and} \bibinfo{person}{Li-Shiuan Peh}.} \bibinfo{year}{2022}\natexlab{}.
\newblock \showarticletitle{{REVAMP: A Systematic Framework for Heterogeneous CGRA Realization}}. In \bibinfo{booktitle}{\emph{Proceedings of the 27th ACM International Conference on Architectural Support for Programming Languages and Operating Systems}}. \bibinfo{publisher}{Association for Computing Machinery}, \bibinfo{address}{New York, NY, USA}, \bibinfo{pages}{918–932}.
\newblock
\urldef\tempurl%
\url{https://doi.org/10.1145/3503222.3507772}
\showDOI{\tempurl}


\bibitem[Carpentieri et~al\mbox{.}(2024)]%
        {Carpentieri2024}
\bibfield{author}{\bibinfo{person}{Nicolo Carpentieri}, \bibinfo{person}{Juan Sapriza}, \bibinfo{person}{Davide Schiavone}, \bibinfo{person}{Daniele Jahier~Pagliari}, \bibinfo{person}{David Atienza}, \bibinfo{person}{Maurizio Martina}, {and} \bibinfo{person}{Alessio Burrello}.} \bibinfo{year}{2024}\natexlab{}.
\newblock \showarticletitle{{Performance evaluation of acceleration of convolutional layers on OpenEdgeCGRA}}. In \bibinfo{booktitle}{\emph{Workshop on Open-Source Hardware}}. ACM, \bibinfo{pages}{1--4}.
\newblock


\bibitem[Das et~al\mbox{.}(2017)]%
        {Das2017}
\bibfield{author}{\bibinfo{person}{Satyajit Das}, \bibinfo{person}{Davide Rossi}, \bibinfo{person}{Kevin Martin}, \bibinfo{person}{Philippe Coussy}, {and} \bibinfo{person}{Luca Benini}.} \bibinfo{year}{2017}\natexlab{}.
\newblock \showarticletitle{{A 142MOPS/mW Integrated Programmable Array accelerator for Smart Visual Processing}}. In \bibinfo{booktitle}{\emph{IEEE International Symposium on Circuits \& Systems}}. \bibinfo{address}{Baltimore, United States}.
\newblock


\bibitem[Duch et~al\mbox{.}(2017)]%
        {Duch227874}
\bibfield{author}{\bibinfo{person}{Loris~Gérard Duch} {et~al\mbox{.}}} \bibinfo{year}{2017}\natexlab{}.
\newblock \showarticletitle{{HEAL-WEAR: an Ultra-Low Power Heterogeneous System for Bio-Signal Analysis}}.
\newblock \bibinfo{journal}{\emph{IEEE Transactions on Circuits and Systems I: Regular Papers}} \bibinfo{volume}{64}, \bibinfo{number}{9} (\bibinfo{year}{2017}), \bibinfo{pages}{14. 2448--2461}.
\newblock
\urldef\tempurl%
\url{https://doi.org/10.1109/Tcsi.2017.2701499}
\showDOI{\tempurl}


\bibitem[Flamand et~al\mbox{.}(2018)]%
        {flamand2018gap}
\bibfield{author}{\bibinfo{person}{Eric Flamand}, \bibinfo{person}{Davide Rossi}, \bibinfo{person}{Francesco Conti}, \bibinfo{person}{Igor Loi}, \bibinfo{person}{Antonio Pullini}, \bibinfo{person}{Florent Rotenberg}, {and} \bibinfo{person}{Luca Benini}.} \bibinfo{year}{2018}\natexlab{}.
\newblock \showarticletitle{{GAP-8: A RISC-V SoC for AI at the Edge of the IoT}}. In \bibinfo{booktitle}{\emph{2018 IEEE 29th International Conference on Application-specific Systems, Architectures and Processors (ASAP)}}. IEEE, \bibinfo{pages}{1--4}.
\newblock


\bibitem[Garofalo et~al\mbox{.}(2020)]%
        {garofalo2020pulp}
\bibfield{author}{\bibinfo{person}{Angelo Garofalo} {et~al\mbox{.}}} \bibinfo{year}{2020}\natexlab{}.
\newblock \showarticletitle{{PULP-NN: Accelerating Quantized Neural Networks on Parallel Ultra-Low-Power RISC-V Processors}}.
\newblock \bibinfo{journal}{\emph{Phil. Trans. of the Royal Society A}} \bibinfo{volume}{378}, \bibinfo{number}{2164} (\bibinfo{year}{2020}), \bibinfo{pages}{20190155}.
\newblock


\bibitem[{GreenWaves Technologies}(2024)]%
        {GreenWaves2024}
\bibfield{author}{\bibinfo{person}{{GreenWaves Technologies}}.} \bibinfo{year}{2024}\natexlab{}.
\newblock \bibinfo{title}{{GAP9}}.
\newblock \bibinfo{howpublished}{\url{https://greenwaves-technologies.com/gap8_mcu_ai/}}.
\newblock


\bibitem[Guthaus et~al\mbox{.}(2001)]%
        {guthaus2001mibench}
\bibfield{author}{\bibinfo{person}{Matthew~R Guthaus}, \bibinfo{person}{Jeffrey~S Ringenberg}, \bibinfo{person}{David Ernst}, \bibinfo{person}{Todd~M Austin}, \bibinfo{person}{Trevor Mudge}, {and} \bibinfo{person}{Richard~B Brown}.} \bibinfo{year}{2001}\natexlab{}.
\newblock \showarticletitle{{MiBench: A free, commercially representative embedded benchmark suite}}. In \bibinfo{booktitle}{\emph{Proceedings of the Fourth IEEE International Workshop on Workload Characterization}}. IEEE, \bibinfo{pages}{3--14}.
\newblock


\bibitem[Heidorn et~al\mbox{.}(2020)]%
        {heidorn2020design}
\bibfield{author}{\bibinfo{person}{Christian Heidorn} {et~al\mbox{.}}} \bibinfo{year}{2020}\natexlab{}.
\newblock \showarticletitle{{Design Space Exploration for Layer-parallel Execution of Convolutional Neural Networks on CGRAs}}. In \bibinfo{booktitle}{\emph{Proc. of the 23th SCOPES}}. \bibinfo{pages}{26--31}.
\newblock


\bibitem[Lee and Lee(2021)]%
        {lee2021specializing}
\bibfield{author}{\bibinfo{person}{Jungi Lee} {and} \bibinfo{person}{Jongeun Lee}.} \bibinfo{year}{2021}\natexlab{}.
\newblock \showarticletitle{{Specializing CGRAs for Light-Weight Convolutional Neural Networks}}.
\newblock \bibinfo{journal}{\emph{IEEE TCAD}} \bibinfo{volume}{41}, \bibinfo{number}{10} (\bibinfo{year}{2021}), \bibinfo{pages}{3387--3399}.
\newblock


\bibitem[Mei et~al\mbox{.}(2003)]%
        {mei2003exploiting}
\bibfield{author}{\bibinfo{person}{Bingfeng Mei} {et~al\mbox{.}}} \bibinfo{year}{2003}\natexlab{}.
\newblock \showarticletitle{{Exploiting Loop-Level Parallelism on Coarse-Grained Reconfigurable Architectures Using Modulo Scheduling}}.
\newblock \bibinfo{journal}{\emph{IEE Proceedings-Computers and Digital Techniques}} \bibinfo{volume}{150}, \bibinfo{number}{5} (\bibinfo{year}{2003}), \bibinfo{pages}{255--261}.
\newblock


\bibitem[Podobas et~al\mbox{.}(2020)]%
        {podobas2020survey}
\bibfield{author}{\bibinfo{person}{Artur Podobas} {et~al\mbox{.}}} \bibinfo{year}{2020}\natexlab{}.
\newblock \showarticletitle{{A Survey on Coarse-Grained Reconfigurable Architectures From a Performance Perspective}}.
\newblock \bibinfo{journal}{\emph{IEEE Access}}  \bibinfo{volume}{8} (\bibinfo{year}{2020}), \bibinfo{pages}{146719--146743}.
\newblock


\bibitem[Tirelli et~al\mbox{.}(2024)]%
        {tirelli2024sat}
\bibfield{author}{\bibinfo{person}{Cristian Tirelli} {et~al\mbox{.}}} \bibinfo{year}{2024}\natexlab{}.
\newblock \showarticletitle{{SAT-based Exact Modulo Scheduling Mapping for Resource-Constrained CGRAs}}.
\newblock \bibinfo{journal}{\emph{ACM Journal on Emerging Technologies in Computing Systems}} (\bibinfo{year}{2024}).
\newblock
\showeprint[arxiv]{2402.12834}


\bibitem[Vazquez et~al\mbox{.}(2024)]%
        {vazquez2024strelastreamingelasticcgra}
\bibfield{author}{\bibinfo{person}{Daniel Vazquez}, \bibinfo{person}{Jose Miranda}, \bibinfo{person}{Alfonso Rodriguez}, \bibinfo{person}{Andres Otero}, \bibinfo{person}{Pascuale~Davide Schiavone}, {and} \bibinfo{person}{David Atienza}.} \bibinfo{year}{2024}\natexlab{}.
\newblock \bibinfo{title}{{STRELA: STReaming ELAstic CGRA Accelerator for Embedded Systems}}.
\newblock
\newblock
\showeprint[arxiv]{2404.12503}~[cs.AR]
\urldef\tempurl%
\url{https://arxiv.org/abs/2404.12503}
\showURL{%
\tempurl}


\bibitem[Wang et~al\mbox{.}(2025)]%
        {wang2025mlir}
\bibfield{author}{\bibinfo{person}{Yuxuan Wang}, \bibinfo{person}{Cristian Tirelli}, \bibinfo{person}{Lara Orlandic}, \bibinfo{person}{Juan Sapriza}, \bibinfo{person}{Rub{\'e}n Rodr{\'\i}guez~{\'A}lvarez}, \bibinfo{person}{Giovanni Ansaloni}, \bibinfo{person}{Laura Pozzi}, {and} \bibinfo{person}{David Atienza~Alonso}.} \bibinfo{year}{2025}\natexlab{}.
\newblock \showarticletitle{An MLIR-based Compilation Framework for CGRA Application Deployment}.
\newblock \bibinfo{journal}{\emph{Applied Reconfigurable Computing. Architectures, Tools, and Applications}} (\bibinfo{year}{2025}).
\newblock


\bibitem[Wijerathne et~al\mbox{.}(2021)]%
        {wijerathne2021himap}
\bibfield{author}{\bibinfo{person}{Dhananjaya Wijerathne} {et~al\mbox{.}}} \bibinfo{year}{2021}\natexlab{}.
\newblock \showarticletitle{{HiMap: Fast and Scalable High-Quality Mapping on CGRA via Hierarchical Abstraction}}.
\newblock \bibinfo{journal}{\emph{IEEE Transactions on Computer-Aided Design of Integrated Circuits and Systems}} \bibinfo{volume}{41}, \bibinfo{number}{10} (\bibinfo{year}{2021}), \bibinfo{pages}{3290--3303}.
\newblock


\bibitem[Wijtvliet et~al\mbox{.}(2021)]%
        {Wijtvliet2021}
\bibfield{author}{\bibinfo{person}{Mark Wijtvliet}, \bibinfo{person}{Henk Corporaal}, {and} \bibinfo{person}{Akash Kumar}.} \bibinfo{year}{2021}\natexlab{}.
\newblock \showarticletitle{{CGRA-EAM—Rapid Energy and Area Estimation for Coarse-grained Reconfigurable Architectures}}.
\newblock \bibinfo{journal}{\emph{ACM Transactions on Reconfigurable Technology and Systems}} \bibinfo{volume}{14}, \bibinfo{number}{4}, Article \bibinfo{articleno}{19} (\bibinfo{date}{Sept.} \bibinfo{year}{2021}), \bibinfo{numpages}{19:1--19:28}~pages.
\newblock
\urldef\tempurl%
\url{https://doi.org/10.1145/3468874}
\showDOI{\tempurl}


\end{thebibliography}

\end{document}